\documentclass[a4paper,10pt]{article}
\usepackage[utf8x]{inputenc}

\title{A Superspace Formulation of the BV 
Action for Higher Derivative Gauge Theories}
\author{ 
Mozzam Khan\\
Department of Mathematics,
King's College London\\
Strand,
London WC2R 2L }
\begin{document}

\maketitle

\begin{abstract}
In this paper we analyse a certain type
 of higher derivative gauge theories which are known to
 possess BRST symmetry associated
 with their higher derivative 
structure. We first show that these theories are also
 invariant under a  anti-BRST symmetry and a double BRST symmetry. 
We then discuss the invariance of these theories under
 shift symmetry in the superspace Batalin Vilkovisky (BV) formalism.  
We show that these theories
 are  manifestly extended BRST invariant. However they are 
 only   extended anti-BRST  invariant  on-shell.
\end{abstract}



\section{Introduction}
Higher derivative gauge theories arise naturally expansion of the  Born-Infeld action for $D$-branes \cite{n1}-\cite{n6}. 
These higher order corrections to gauge theories become relevant for studding $D$-branes beyond the slowly varying field approximation. 
The expansion of the action can be viewed as an expansion in  $\alpha '$ and 
the Yang-Mills coupling $g$, with $\alpha ' g F_{\mu\nu}$ being dimensionless. 
Thus the lowest order  term are corrected by terms like $\alpha '^2 g^2 F^{\mu\nu}F_{\mu\nu} F^{\rho\sigma}F_{\rho\sigma}$ and
 $\alpha '^4 g^4 F^{\mu\nu}F_{\mu\nu} F^{\rho\sigma}F_{\rho\sigma}F^{\tau\lambda}F_{\tau\lambda}$  along with terms like 
$\alpha '^3 g^2 D^2 F^{\mu\nu}F_{\mu\nu} F^{\rho\sigma}F_{\rho\sigma}$
 and $\alpha '^4 g^2 D^4 F^{\mu\nu}F_{\mu\nu} F^{\rho\sigma}F_{\rho\sigma}$.
 
Higher derivative gauge theories are also important in the study of linearly polarized laser beam propagating through transverse 
magnetic field \cite{la1}. It is hoped that induced ellipticity of the laser beam  can be explained by 
effective cubic order Lagrangian  in the field strength \cite{la2}.

However existence of higher derivatives  causes theses 
theories to possess 
negative norm states or ghost states associated 
with their higher derivative structure \cite{3a}-\cite{4a}. 
  It is well know how to deal with the problems associated 
with negative norm states induced by the Faddeev-Popov ghosts  in 
conventional gauge theories  by  means of BRST and anti-BRST
 symmetries  \cite{7a}-\cite{8a}. In fact,  gauge invariance in gauge theories
 also has  very interesting consequences in string theory \cite{za}-\cite{zb}. So, the BRST symmetry has been studied in string theory 
\cite{za1}-\cite{zb1}. 
 This BRST or anti-BRST invariance of the sum of the classical Lagrangian, the gauge fixing term and the ghost term can be used to
  remove all the negative norm states associated with the  Faddeev-Popov ghosts.

Recently BRST symmetry in higher derivative  gauge theories has been studied \cite{9a}.  This BRST symmetry is not due to gauge
fixing but is an intrinsic feature of a certain type of higher derivative gauge theories.  
In fact it was demonstrated in   \cite{9a}-\cite{faizalmo} that even certain scalar field theories possess BRST and anti-BRST symmetries along with 
double BRST symmetry. Furthermore, the effect of shift symmetry on these scalar theories has also been recently analysed  in the 
superspace BV-formalism \cite{faizalmo}. BV formalism  \cite{10a}-\cite{11a} in the context of both 
the extended BRST, and the extended anti-BRST symmetries of Faddeev-Popov ghosts \cite{12a}-\cite{13a} along  
 a superspace formalism for it is also well understood \cite{14a}-\cite{15a}. 

In this paper we will analyse 
certain gauge theories known to possess a BRST symmetry  with their higher 
derivative structure \cite{9a} on similar line to what was done in \cite{faizalmo}. So we will first  show that 
these higher derivative 
gauge theories also   possesses anti-BRST symmetry along with   the double BRST symmetry. 
Than we shall study the effect of shift symmetry on these theories in the 
Batalin Vilkovisky (BV) formalism. 
So we will study these higher derivative gauge theories in the superspace BV 
formalism in analogy to what  was done for scalar field theories BRST symmetry associated  with their higher derivative structure. 

\section{BRST and Anti-BRST Invariant Lagrangian Density}
The Lagrangian density $\mathcal{L}$ of  higher derivative gauge  theory that we will analyze in this paper is 
\begin{equation}
\mathcal{L}= Tr \left[ \frac{1}{2}\mathcal{O} A ^\mu \mathcal{O} A _\mu + \overline{c}^\mu\mathcal{O} c_\mu \right], \label{2a}
\end{equation}
where $c_\mu$ is a ghost field, $\overline{c}_\mu$  is an anti-ghost field, $ A _\mu$ is a gauge field  and $\mathcal{O}$ 
depends on the order of the theory. 
The Lagrangian density  given in Eq. $(\ref{2a})$ can then be rewritten using an auxiliary field $L_\mu$ as
\begin{equation}
\mathcal{L}=Tr \left[L^\mu\mathcal{O} A _{\mu}-\frac{1}{2}L^\mu L_\mu + \overline{c}^\mu\mathcal{O} c_\mu \right]. \label{4a}
\end{equation}

We can also write this Lagrangian density   explicitly for any order of the theory. For  example for  a fourth order theory it is given by
\begin{equation}
Tr \left[L^\mu \mathcal{O}A_{\mu} \right] = Tr  \left[L^\mu \left(D^\nu F_{\mu \nu} 
+ \frac{1}{2\alpha}\partial_\mu\partial^\nu A_\nu\right) \right].\label{3a}
\end{equation} 
With this the  Lagrangian density  for a fourth order theory can be explicitly written as 
\begin{eqnarray}
\mathcal{L}&=&\frac{1}{2} Tr\left[D^\nu F_{\mu \nu} D_\rho F^{\rho \mu} +
 \frac{1}{4 \alpha^{2}}\partial^\mu \partial^\rho  A_\rho \partial_{\mu}.\partial^\sigma A_\sigma \right.\\ \nonumber 
&&+\frac{1}{\alpha} D^{\nu}F_{\nu \mu} \partial^{\mu} \partial^\rho A_\rho + 
\overline{\mathcal{F}}^{\mu \nu}\mathcal{F}_{\mu \nu}-2i \{\overline{c}^{\mu},c^{\nu}\}F_{\nu \mu} 
\\ \nonumber
&&\left. +\frac{1}{\alpha}\partial^{\mu}\overline{c}_{\mu}\partial^{\nu}c_{\nu}\right].
\end{eqnarray}
However we will use the Lagrangian density given by Eq. $(\ref{4a})$ in this paper so as to keep the results very general
 and applicable to any  order gauge theory. 
So we will not define the exact form of $\mathcal{O}$ in this paper as it will in general depend on the order of the theory. 

The  Lagrangian density given by Eq $(\ref{4a})$ is known to possess a BRST symmetry \cite{9a}, as it is  invariant under the following BRST transformations
\begin{eqnarray}
\delta  A _\mu&=&c_\mu,\nonumber\\
\delta \overline{c}_\mu&=&-L_\mu,\nonumber\\
\delta c_\mu&=&0,\nonumber\\
\delta L_\mu&=&0.
\end{eqnarray}
We note that this  Lagrangian density given by Eq $(\ref{4a})$ is also invariant under the following anti-BRST transformations 
\begin{eqnarray}
\overline{\delta} A _\mu&=&\overline{c}_\mu,\nonumber \\
\overline{\delta c}_\mu&=&0,\nonumber \\
\overline{\delta}c_\mu&=&L_\mu, \nonumber \\
\overline{\delta}L_\mu&=&0,\label{6a}
\end{eqnarray}
Now we can write this Lagrangian density as a total BRST variation or a total anti-BRST variation as follows: 
\begin{eqnarray}
\mathcal{L}&=&\overline{\delta} Tr\left[ \left( c^\mu\left(\mathcal{O} A _\mu -\frac{1}{2}L_\mu \right) \right) \right] \nonumber \\
&=&-\delta Tr\left[ \left(\overline{c}^\mu\left(\mathcal{O} A _\mu-\frac{1}{2}L_\mu \right) \right) \right]. \label{7aab}
\end{eqnarray}
However it can also be written as a total double BRST variation. This is a feature of gauge theories with higher derivative structure
as in conventional gauge theories this can only be done in Landau gauge or in non-linear gauges. 
\begin{eqnarray}
\mathcal{L}
&=&\frac{1}{2}\overline{\delta}\delta Tr\left[ \left( A ^\mu \mathcal{O} A _\mu-c^\mu\overline{c}_\mu \right) \right] \nonumber \\
&=&-\frac{1}{2}\delta\overline{\delta} Tr\left[ \left( A ^\mu \mathcal{O} A _\mu - c^\mu\overline{c}_\mu \right) \right]. 
\end{eqnarray}  
\section{Extended BRST and Anti-BRST Lagrangian Density}
The extended BRST  and anti-BRST invariant Lagrangian density is obtained by  
 requiring the Lagrangian density  to be invariant under  the original BRST transformations, the original anti-BRST 
transformations  and the shift
transformations of the original fields given by 
\begin{eqnarray}
 A _\mu &\to&  A _\mu - \tilde{ A }_\mu, \nonumber \\
c_\mu &\to& c_\mu - \tilde{c}_\mu, \nonumber \\
\overline{c}_\mu &\to& \overline{c}_\mu - \tilde{\overline{c}}_\mu,\nonumber\\
L_\mu &\to&  L_\mu - \tilde{ L}_\mu.
\end{eqnarray}
Thus the extended BRST invariant Lagrangian density given 
 is invariant under the following  extended BRST symmetry with the transformations
\begin{eqnarray}
\delta A _\mu=\psi_\mu, && \delta \tilde A _\mu=(\psi_\mu- (c_\mu- \tilde c_\mu)),\nonumber \\ 
\delta c_\mu =\epsilon_\mu, && \delta \tilde{c}_\mu=\epsilon_\mu,\nonumber \\
\delta \overline c_\mu = \epsilon_\mu,&&  \delta \tilde{\overline c}_\mu=(\overline \epsilon_\mu + (L_\mu-\tilde L_\mu)), \nonumber\\
\delta L_\mu =\rho_\mu, && \delta \tilde L_\mu=\rho_\mu.
\end{eqnarray}
Here, $\psi_\mu, \epsilon_\mu, \overline \epsilon_\mu $ and $\rho_\mu$ are the ghost fields associated with the shift symmetries of the original fields
 $ A _\mu, c_\mu,
 \overline {c}_\mu$ and $L_\mu$ respectively. Following the standard BV-formalism we also  
add anti-fields with opposite parity to the original fields along with  new 
auxiliary fields. These  anti-fields transform into new 
auxiliary fields under BRST transformations,
\begin{eqnarray}
\delta  A ^{*}_\mu&=&-b_\mu,\nonumber\\
\delta c^{*}_\mu&=&-B_\mu,\nonumber\\
\delta \overline c^{*}_\mu&=&-\overline{B}_\mu, \nonumber \\
\delta L^{*}_\mu&=&-\overline{b}_\mu.
\end{eqnarray}
The   BRST transformations of these ghosts associated with the shift symmetry and these new auxiliary fields vanish 
\begin{eqnarray}
\delta \psi_\mu = 0,&&\delta b_\mu=0,\nonumber\\
\delta \epsilon_\mu =0,&& \delta B_\mu=0,\nonumber \\
\delta \tilde{\epsilon}_\mu=0,&&\delta \overline B_\mu=0,\nonumber\\
\delta \rho_\mu=0, && \delta \overline{b}_\mu=0.
\end{eqnarray}

The original 
and shifted fields obey the extended anti-BRST transformations,
\begin{eqnarray}
\overline{\delta}\tilde{ A }_\mu= A ^{*}_\mu, && \overline{\delta} A _\mu=  A ^{*}_\mu+(c_\mu-\tilde{\overline{c}}_\mu), \nonumber\\
\overline{\delta}\tilde{c}_\mu=\overline{c}^{*}_\mu, && \overline{\delta}c_\mu= \overline c^{*}_\mu+(L_\mu-\overline{L}_\mu), \nonumber\\
\overline{\delta}\tilde{\overline{c}}_\mu={c}^{*}_\mu, && \overline{\delta}\overline{c}_\mu={c}^{*}_\mu,\nonumber\\
\overline{\delta}\tilde{L}_\mu=L^{*}_\mu, && \overline{\delta}L_\mu= L^{*}_\mu.
\end{eqnarray}
The ghost fields associated with the shift symmetry have the following extended anti-BRST transformations,
\begin{eqnarray}
\overline{\delta}\psi_\mu&=&b_\mu+(L_\mu-\tilde{L}_\mu),\nonumber\\
\overline{\delta} \epsilon_\mu&=&B_\mu,\nonumber\\
\overline{\delta}\overline{\epsilon}_\mu&=&\overline{B}_\mu,\nonumber\\
\overline{\delta} \rho_\mu&=&\overline{b}_\mu,
\end{eqnarray}
and the extended anti-BRST transformations of the anti-fields of the auxiliary fields associated with the shift symmetry vanishes,
\begin{eqnarray}
\overline{\delta }b_\mu =0, && \overline{\delta}  A ^{*}_\mu =0,\nonumber\\
\overline{\delta} B_\mu=0, &&\overline{\delta} c^{*}_\mu=0,\nonumber\\
\overline{\delta} \overline{B}_\mu=0, &&\overline{\delta} \overline c^{*}_\mu=0,\nonumber\\
\overline{\delta }\overline{b}_\mu=0, && \overline{\delta} L^{*}_\mu=0.
\end{eqnarray}

Now we can find a  Lagrangian density which is invariant under the extended BRST transformations and also invariant 
under the  extended anti-BRST transformation
 at least on-shell. It has to be invariant under extended anti-BRST transformation on-shell as  the extended anti-BRST transformation
 reduce to reduce to standard anti-BRST transformations on-shell.

Following what was done for scalar field theories in \cite{faizalmo}, we choose the  
Lagrangian density in such a way that it fixes the shift symmetry in such a way that the tilde fields will be made to
 vanish. Thus the  Lagrangian density with extended BRST symmetry and on-shell extended anti-BRST symmetry is given can be written as:   
\begin{eqnarray}
\mathcal{L}_{\rm{tot}}
&=& Tr\left[ A ^{*\mu} c_\mu-c^{* \mu}L_\mu-\left( A ^*_{\mu} +\frac{\delta \Psi}{\delta L^\mu}\right)\psi^\mu \right. \nonumber \\
&& \left.+\left(\overline{c}^{*}_\mu + \frac{\delta \Psi}{\delta c^\mu}\right)\epsilon^\mu-\left(c^{*\mu} - \frac{\delta \Psi}{\delta \overline{c}^\mu}\right)
\overline{\epsilon}^\mu+\left(L^{*}_\mu-
 \frac{\delta \Psi}{\delta L^\mu}\right)\rho^\mu \right],
\end{eqnarray}
with 
\begin{equation}
 \Psi=-Tr[\overline{c}^\mu(\mathcal{O} A _\mu-L_\mu/2)].
\end{equation}

Now we can obtain  explicit expression for the anti-fields in terms of the original fields 
 by  integrating out the ghosts associated with the shift symmetry, 
\begin{eqnarray}
 A _{\mu}^*&=&-\frac{\delta\Psi}{\delta A ^\mu},\nonumber\\
\overline{c}_{\mu}^*&=&-\frac{\delta \Psi}{\delta c^\mu},\nonumber\\
c^{*}_\mu&=&\frac{\delta \Psi}{\delta \overline{c}^\mu},\nonumber\\
L^{*}_\mu&=&\frac{\delta \Psi}{\delta L^\mu}.
\end{eqnarray}
Now we can compare this expression with the original one and that will lead to the 
identification of these anti-fields with the original fields as follows:
\begin{eqnarray}
 A _{\mu}^*&=& \mathcal{O}\overline{c}_\mu,\nonumber\\
\overline{c}_{\mu}^*&=&0,\nonumber\\
c_{\mu}^*&=&-\mathcal{O} A _\mu+ \frac{L_\mu}{2},\nonumber\\
L_{\mu}^*&=&\frac{\overline{c}_\mu}{2}. \label{7b}
\end{eqnarray} 
\section{Extended  Superspace }
This Lagrangian density can be given a geometric structure in terms of superspace formalism. 
However as the ghosts are vector field for these theories we will need to define superfield with free Lorentz index
 with  two anti-commutating
parameters, namely $\theta$ and $\overline{\theta}$. Thus we define the following superfields:
\begin{eqnarray}
\varphi_\mu(x,\theta,\overline{\theta})&=& A _\mu+\theta \psi_\mu+\overline{\theta}( A ^{*}_\mu+(\overline{c}_\mu-\tilde{\overline{c}}_\mu))+\theta \overline{\theta}(b_\mu+ (L_\mu-\tilde{L}_\mu)),
\nonumber\\
\tilde{\varphi}_\mu(x,\theta,\overline{\theta})&=&\tilde{ A }_\mu+\theta(\psi_\mu-(c_\mu-\tilde{c}_\mu))+\overline{\theta} A ^{*}_\mu+\theta \overline{\theta}b_\mu,
\nonumber\\
\chi_\mu(x,\theta,\overline{\theta})&=&c_\mu+\theta \epsilon_\mu
+\overline{\theta}(c^{*}_\mu+(L_\mu-\tilde{L}_\mu))+\theta\overline{\theta}B_\mu,
 \nonumber\\
\tilde{\chi}_\mu(x,\theta,\overline{\theta})&=&\tilde{c}_\mu+\theta \epsilon_\mu+\overline{\theta}c^{*}_\mu+\theta\overline{\theta}B_\mu,
\nonumber\\
\overline{\chi}_\mu(x,\theta,\overline{\theta})&=&\overline{c}_\mu+\theta\overline{\epsilon}_\mu+\overline{\theta} \overline{c}^{*}_\mu +\theta \overline{\theta}\overline{B}_\mu,
\nonumber\\
\tilde{\overline{\chi}}_\mu(x,\theta,\overline{\theta})&=&\tilde{\overline{c}}_\mu+\theta(\overline{\epsilon}_\mu+(L_\mu-\tilde{L}_\mu)) + \overline{\theta}\overline{c}^{*}_\mu
+\theta \overline{\theta}\overline{B}_\mu. \label{7c}
\end{eqnarray}
Now we can define $ \tilde{\mathcal{L}}$ as the Lagrangian density that depends only on the shifted fields. 
Thus from the above expressions, we get 
\begin{eqnarray}
 \tilde{\mathcal{L}}&=&\frac{\partial}{\partial \overline{\theta}}\frac{\partial}{\partial \theta}
 Tr \left[-\frac{1}{2}\tilde{\varphi}^\mu\tilde{\varphi}_\mu+\tilde{\chi}^\mu\tilde{\overline{\chi}}_\mu \right ]\nonumber \\ &=&
-Tr \left [ b^\mu\tilde{ A }_\mu- A ^{*\mu}(\psi_\mu-(c_\mu-\tilde{c}_\mu))-\overline{B}^\mu\tilde{c}_\mu+\overline{c}^{*\mu}\epsilon_\mu \right. \\ && \left. +B^\mu\tilde{\overline{c}}_\mu-c^{*\mu}(\overline{\epsilon}_\mu
+(L_\mu-\tilde{L}_\mu)) \right ].
\end{eqnarray}
Being the $\theta \overline{\theta}$ component of a superfield, 
this gauge-fixing Lagrangian density is manifestly invariant under extended BRST and anti-BRST
 transformations. 

 We now define  $\mathcal{L}$ 
as the Lagrangian density for the original fields and write it as 
\begin{equation}
\mathcal{L}=\frac{\partial}{\partial \theta}(\delta(\overline{\theta}) \Phi(x, \theta, \overline{\theta})), \label{7d}
\end{equation}
where 
\begin{equation}
\Phi(x,\theta,\overline{\theta})= Tr[\varphi^\mu(x,\theta,\overline{\theta})\chi_\mu(x,\theta,\overline{\theta})].
\end{equation}
We can thus express it as follows:
\begin{equation}
\Phi(x,\theta,\overline{\theta})= \Psi+\theta \delta \Psi+ \overline{\theta}\overline{\delta}\Psi
+\theta \overline{\theta} \delta \overline{\delta}\Psi. \label{3d}
\end{equation}
The component of $\theta \overline{\theta}$ can be made to vanish on-shell.
This Lagrangian density is  manifestly invariant under extended BRST transformations. It is
 also invariant under  on-shell extended anti-BRST transformations.  

The complete Lagrangian density is given by the sum of the above two Lagrangian densities. It can therefore be written as: 
\begin{eqnarray}
\mathcal{L}_{\rm{tot}}&=&\tilde{\mathcal{L}} +\mathcal{L}\nonumber\\
&=&\frac{\partial}{\partial \overline{\theta}} \frac{\partial}{\partial \theta} Tr \left [\left(-\frac{1}{2}\tilde{\varphi}^\mu\tilde{\varphi}_\mu+
 \tilde{\chi}^\mu\tilde{\overline{\chi}}_\mu \right) + \frac{\partial}{\partial \theta}(\delta(\overline{\theta})\Phi(x,\theta,\overline{\theta})) \right ]\nonumber\\
&=&- Tr \left[ b^\mu\tilde{ A }_\mu-\overline{B}^\mu\tilde{c}_\mu+B^\mu\tilde{\overline{c}}_\mu - ( A ^{*}_\mu+\frac{\delta \Psi}{\delta  A ^\mu})\psi^\mu
 \right. \nonumber \\ && \left. + A ^{*\mu}(c_\mu-\tilde{c}_\mu)-c^{*\mu}(L_\mu-\tilde{L}_\mu) \right. \nonumber\\
&& \left. +(\overline{c}^{*\mu}
 + \frac{\delta \Psi_\mu}{\delta c^\mu})\epsilon^\mu-(c^{*^\mu}-
\frac{\delta \Psi}{\delta \overline{c}^\mu})\overline{\epsilon}^\mu \right]. \label{8d}
\end{eqnarray}
We can redefine the auxiliary field as $L_\mu-\tilde{L}_\mu\to L_\mu$ because  the 
 combination $(L_\mu+\tilde{L}_\mu)$ can then be integrated out and absorbed into the normalization constant.

\section{Conclusion}
We have analysed higher derivative gauge  theories with suitably added ghost terms in the superspace BV formalism.
These theories are expressed as  a total double BRST variation. We have also analysed the effect of shift symmetry on these theories. 
 These theories  are  invariant under 
extended BRST transformations but they are only invariant under on-shell extended   anti-BRST  
transformations.

It will  be interesting to develop a supersymmetric version of this theory and apply it to higher derivative super-Yang-Mills theories.
 However so far no 
higher derivative super-Yang-Mills theory is know to possess a BRST symmetry associated with its higher derivative structure. 
But we can proceed to construct 
such a theory by considering the theory studied here as the bosonic part of the higher derivative super-Yang-Mills theory.  
It will also be interesting to apply the
 results of this paper to higher derivative Chern-Simons theories.
It will be  interesting to generalise the results of this paper to 
 de Sitter and anti-de Sitter spacetimes \cite{desitter}-\cite{desitter1}.
It will also be interesting to investigate higher derivatives in gauge theory using 
 Wheeler-DeWitt 
equation \cite{zp1}-\cite{zq}.

\end{document}